\documentclass[a4paper]{article}
\usepackage[margin=1in]{geometry}
\usepackage{booktabs}
\usepackage{authblk}
\usepackage{amsmath}
\usepackage{csvsimple}
\usepackage{multirow}
\usepackage{caption}
\usepackage{subcaption}
\usepackage{rotating}
\usepackage{tikz}
\usepackage{graphicx}
\usepackage{anyfontsize}
\usepackage{titlesec}
\titleformat{\chapter}[display]{\huge\bfseries}{\chaptertitlename~\thechapter}{20pt}{\Huge}
\titlespacing{\chapter}{0pt}{50pt}{40pt}

\title{
Photon Masks for the ILC Positron Source with 175 and 250 GeV Electron Drive Beam}
\author{K. Alharbi \textsuperscript{1} \textsuperscript{2} \textsuperscript{3},  S. Riemann \textsuperscript{3}, A. Alrashdi \textsuperscript{1}, G. Moortgat-Pick \textsuperscript{2} \textsuperscript{4}, A. Ushakov \textsuperscript{2}, P. Sievers \textsuperscript{5}\\
 King Abdul-Aziz City for Science and Technology (KACST), Riyadh, Saudi Arabia \textsuperscript{1}\\
University of Hamburg, Hamburg, Germany \textsuperscript{2}\\
Deutsches Elektronen-Synchrotron (DESY), Zeuthen, Germany \textsuperscript{3}\\
 Deutsches Elektronen-Synchrotron (DESY), Hamburg, Germany \textsuperscript{4}\\
CERN, Geneva 23, Switzerland \textsuperscript{5}}

\date{}
\begin{document}
\begin{flushright}

\hfill
DESY 21-084

  \end{flushright}
\begin{minipage}{\textwidth}
   \maketitle
   \thanks

\begin{abstract}

\fontsize{13}{13}\selectfont

In the future the International Linear Collider (ILC), a helical undulator-based polarized positron source, is expected to be chosen. A high energy electron beam passes through a superconducting helical undulator in order to create circularly polarized photons which will be directed to a conversion target, resulting in electron-positron pairs. The resulting positron beam is longitudinally polarized. Since the photons are produced with an opening angle and pass through a long superconducting helical undulator, some of these photons will strike the undulator walls. Therefore photon masks must be placed along the undulator line in order to keep the power deposited in the undulator walls below the acceptable limit of 1W/m. The baseline design of the ILC is focused on 250 GeV center-of-mass energy and upgrade to center-of-mass energies of 350 and 500 GeV is foreseen. This paper shows a detailed study of the ideal power deposited along the masks for both 350 and 500 GeV center-of-mass energies.

\end{abstract}

\end{minipage}

\begin{flushleft}
\rule{200pt}{0.5pt}\\
\fontsize{10}{10}\selectfont
{"Talk presented at the International Workshop on Future Linear Colliders (LCWS2021), 15-18 March 2021. C21-03-15.1."\\
ksalharbi@kacst.edu.sa}
\end{flushleft}

\newpage

\fontsize{12}{22}\selectfont
\section{Introduction}

Nowadays, physicists in the field of elementary particle physics are interested in discovering new particles at hundreds or thousands of GeV center-of-mass energies. One of the most significant current projects in this field is the International Linear Collider (ILC). The proposed ILC would collide positrons with electrons using center-of-mass energy of 250-500 GeV (extendable to 1 TeV). 

The Technical Design Report (TDR) \cite{adolphsen2013international} of the ILC describes two possible positron sources; namely, the helical undulator and an e-driven source. The comparison between these two sources has been studied in \cite{TDR2018}. One of the most vital benefits of using a helical undulator as a baseline for the ILC is the possibility of producing longitudinally polarized positrons \cite{moortgat2008polarized}. 

\begin{figure}[h]
\centering
\includegraphics[scale=0.39]{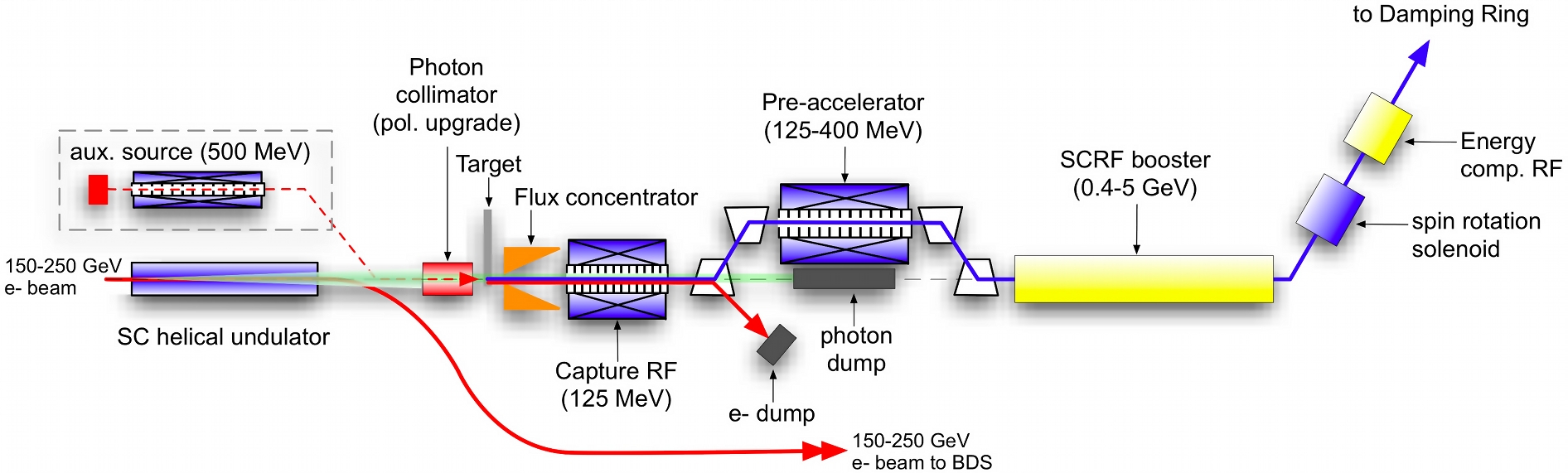}
\caption{Schematic layout of the ILC Positron Source \cite{adolphsen2013international}.}
\label{fig:ILC}
\end{figure}

The helical undulator can produce a high-intensity polarized positron beam. Figure \ref{fig:ILC} shows the schematic layout of positron-beam production for the ILC. In the ILC, circularly polarized photons with multi-MeV are generated when a multi-GeV electron beam passes through a helical undulator. These photons are then directed to a thin metal target to produce electron-positron pairs. The electron-positron pair will inherit the photon polarization ($P_\gamma$) \cite {flottmann1993investigations}. After that, the photons and electrons will be dumped, while the longitudinally polarized positrons will be captured and accelerated up to 5 GeV before being sent to the damping ring (DR).

The general layout of the helical undulator has been described in the ILC TDR \cite {adolphsen2013international}. A pair of helical undulator modules with a length of 1.75 m for each is installed in a cryomodule which is a 4.1 m long. The helical undulator module has an aperture and period of 5.85 and 11.5 mm, respectively. To steer and focus the electron beam through the helical undulator, 23 quadrupoles are placed along the helical undulator. Each quadrupole is placed after each three cryomodules. In the ILC-250GeV, 66 cryomodules are needed to produce the required positrons (1.5 e$+$/e$-$). The needed area for the helical undulator is 319.828 m, giving 231 m the total active length.

\section{The ILC positron source parameters}

The ILC positron source parameters had been updated since the start of the ILC project in 2005. Following the discovery of the Higgs boson, which has a mass of 125 GeV, a 250 GeV center-of-mass energy was chosen for the first stage in the ILC. The later stages are to upgrade to 350 and 500 GeV center-of-mass-energies for studying the top quark threshold and exciting processes of the Standard Module and beyond, respectively.

When the center-of-mass energy is changed, the other parameters of the helical undulator must be changed in order to produce the required positrons (1.5 e$+$/e$-$). Table \ref{table1} shows the parameters for the baseline ILC Positron Source and table \ref{table2} shows the parameters required for 350 and 500 GeV center-of-mass-energies.

\begin{table}[h]

\caption[ILC undulator parameters]{The parameters for the ILC-250 GeV Positron Source.}
\centering % used for centering table
\begin{tabular}{l*{6}{l}r} % centered columns (4 columns)
\hline\hline %inserts double horizontal lines
Parameters & Values & Unit \\ [0.5ex] % inserts table

\hline % inserts single horizontal line
Centre-of-mass energy  & 250 & GeV  \\ % inserting body of the table
Undulator period & 11.5 & mm\\
Undulator K & 0.85 \\
Electron number per bunch & $2\times10^{10}$ \\
Number of bunches per pulse & 1312 \\
Pulse rate &  5.0 & Hz \\
Cryomodule Length & 4.1&  m \\ 
Effective magnet length & 3.5 & m \\
Undulator aperture & 5.85 &mm \\
Number of quadrupoles  & 23 \\ 
Quadrupole spacing & 14.538 & m\\
Quadrupole length & 1&  m \\
Total active undulator length & 231&  m \\
Total lattice length & 319.828 & m \\

\hline %inserts single line
\end{tabular}
\label{table1} % is used to refer this table in the text
\end{table}

\begin{table}[h]

\caption[ILC undulator parameters]{Positron Source Parameters for 350 GeV and 500 GeV center-of-mass-energies.}
\centering % used for centering table
\begin{tabular}{l*{6}{l}r} % centered columns (4 columns)
\hline\hline %inserts double horizontal lines
Parameters & Values \\ [0.5ex] % inserts table

\hline % inserts single horizontal line
Centre-of-mass energy  (GeV)  & 350 & 500 & GeV \\ % inserting body of the table
Undulator period & \multicolumn{2}{c}{11.5}& mm\\
Required undulator field & 0.698 & 0.42 &\\
Undulator K & 0.75 & 0.45 &\\
Electron number per bunch &  \multicolumn{2}{c} {$2\times10^{10}$} &\\
Number of bunches per pulse &  \multicolumn{2}{c} {1312}& \\
Pulse rate &  \multicolumn{2}{c} { 5.0} &Hz \\
Cryomodule Length & \multicolumn{2}{c} { 4.1} &m \\

Effective magnet length & \multicolumn{2}{c} {3.5}& m \\
Undulator aperture & \multicolumn{2}{c} { 5.85} &mm \\
Quadrupole spacing &  \multicolumn{2}{c} {14.538 }&m\\
Quadrupole length &  \multicolumn{2}{c} {1} &m \\
Total active undulator length &  \multicolumn{2}{c} {147}& m \\

\hline %inserts single line
\end{tabular}
\label{table2} % is used to refer this table in the text
\end{table}

\section{The ILC Helical Undulator Photon Masks}

In the ILC a long superconducting helical undulator is used. Since the photons in the helical undulator are produced with an opening angle which is determined by the energy of the electron beam (which is proportional to 1/$\gamma$) the undulator walls will be bombed by some of these photons. To achieve the required vacuum pressure, the power deposited on the undulator walls due to the synchrotron radiation must be below 1 W/m \cite {scott2008investigation}. This was studied for ILC-250GeV on \cite {alharbi2020energy}. This study shows that 22 photon masks with 0.22 cm radius must be placed along the undulator line to keep the power deposited at the undulator walls below the limit. Each photon mask is placed behind each quadrupole.

\subsection{Photon Mask Design for The ILC Helical Undulator}
The photon mask for the ILC helical undulator must be designed in order not only to stop the primary incident photons but also to stop the secondary particles. A possible photon mask design with a high photon absorption efficiency for the ILC helical undulator was proposed for ILC-250GeV in \cite {IPAC21}. 

\begin{figure}[h]
\centering
\includegraphics[scale=0.6]{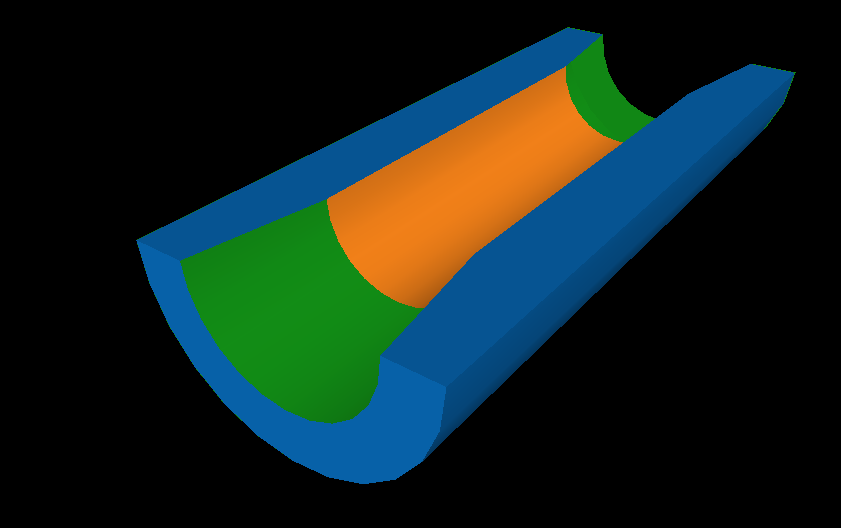}
\caption{A longitudinal section of the photon mask.}
\label{fig:LCWS21Mask}
\end{figure}

\begin{figure}[h]
\centering
\includegraphics[scale=2.5]{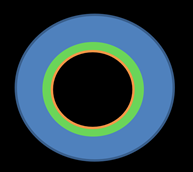}
\caption{A transverse section of the beam pipe showing the photon mask in the beam view.}
\label{fig:LCWS21MaskHole1}
\end{figure}

The photon mask has a cylindrical geometry. Figure \ref{fig:LCWS21Mask} shows a longitudinal section of the photon mask and  figure \ref{fig:LCWS21MaskHole1} shows the transverse section of the beam pipe showing the photon mask in the beam view. Firstly, one must choose the dimensions of the photon mask. Since the saved area for the helical undulator is limited \cite{TDR2018}, the most extended length of the photon mask would be 30 cm. The photon mask outer diameter is 15 cm.

To reduce the wake-fields, the mask should be tapered. The inner radius of the mask is tapered in the first and last 5 cm of the mask length as shows by green areas in figure \ref{fig:LCWS21Mask}. In the first 5 cm, the inner radius is tapered from 0.2925 cm to 0.22 cm, and the inner radius in the last 5 cm is tapered from 0.22 cm to 0.2925 cm, which is the beam pipe radius. The inner radius of the photon mask between these tapered sections is 0.22 cm, as shown by orange area in figure \ref{fig:LCWS21Mask}.

Secondly, one must select the appropriate material for the photon mask. With such a short photon mask, and in order to stop photons with incident average energy in MeV range, the mask material should have small radiation length, high atomic number and high density. Copper is one possible material which can be used for the photon mask. Table \ref{copper} summarizes the properties of copper.

\begin{table}[h]

\caption[ILC undulator parameters]{Properties of the copper.}
\centering % used for centering table
\begin{tabular}{l*{6}{l}r} % centered columns (4 columns)
\hline\hline %inserts double horizontal lines

Parameter &              Unit  & Copper \\

\hline % inserts single horizontal line

Atomic Number && 29  \\
Density & g/$cm^3$  & 8.96     \\
Thermal Conductivity   &  W/(m.K)& 401            \\
Heat Capacity &  J/g/K  &  0.385    \\
Melting Point &  K  & 1357.77 \\
Radiation Length & cm  & 1.436  \\

\hline %inserts single line
\end{tabular}
\label{copper} % is used to refer this table in the text
\end{table}

\subsection{Ideal Power Deposited at Mask} 

The power deposited at the photon masks by the ILC-250 GeV center-of-mass energy using the aforementioned photon mask design has been studied in \cite {IPAC21}. Since the later stages in the ILC are to upgrade to 350 and 500 GeV center-of-mass-energies, the power deposited at the photon masks by the 350 and 500 GeV center-of-mass energies should be studied. 

The photon spectra from the helical undulator are simulated by the Helical Undulator Synchrotron Radiation (HUSR) code \cite{newton2010rapid} and \cite {newton2010modeling}. It is written in C++. Besides the photon spectra, HUSR can calculate the $P_\gamma$.

Table \ref{table2} shows the required parameters for 350 and 500 GeV center-of-mass energies to produce the required positrons (1.5 e$+$/e$-$). There will be 66 cryomodules in the ILC helical undulator, giving a 231 m active length. In the case of 350 and 500 GeV center-mass energies only 42 cryomodules are needed.

\begin{table}[h]

\caption[ILC undulator parameters]{Power deposited at the photon mask for 350 and 500 GeV Center-Of-Mass energies.}
\centering % used for centering table
\begin{tabular}{l*{6}{l}r} % centered columns (4 columns)
\hline\hline %inserts double horizontal lines

Parameter &              Unit  & 350 GeV & 500 GeV\\

\hline % inserts single horizontal line

Average Incident Photon Energy & MeV&   2.54   & 1.75  \\
Incident Photon Number & $\times10^{14}$/sec & 4.6  & 0.74\\
Power Deposited & W  &  186 &  21  \\

\hline %inserts single line
\end{tabular}
\label{HUSR} % is used to refer this table in the text
\end{table}

For 350 and 500 GeV center-of-mass energies, the worst result for both the energy deposited at the masks and power distribution at the target is produced by using the first 42 cyromodules. In contrast, the best result is produced by using the last 42 cryomodules. In order to study the worst scenario of the power deposited at photon masks, the first 42 cryomodules are chosen to be used for both center-mass energies in this paper. In addition, since the photons in the helical undulator are produced with an opening angle, it is clear that the last photon mask is receiving the highest amount of the power. Therefore, only the ideal power deposited at the last mask is studied here, taking into account the power deposited in the previous masks and the electron loss energies along the undulator line which are $\simeq$ 2.6 GeV and  $\simeq$ 2 GeV for 350 GeV and 500 GeV center-of-mass energies, respectively.

Table \ref {HUSR} shows the average incident photon energy, incident photon number and power deposited at the mask for both 350 and 500 GeV center-of-mass energies.

\subsection{Energy Deposited Along The Photon Mask}

FLUKA Monte Carlo code for particle tracking and particle interactions with matter \cite{fluka} was used to simulate the energy deposited along the photon mask for both 350 and 500 GeV center-of-mass energies. The incident photons simulated by HUSR were used as input to FLUKA. 

\begin{figure}[h]
\centering
\includegraphics[scale=1.0]{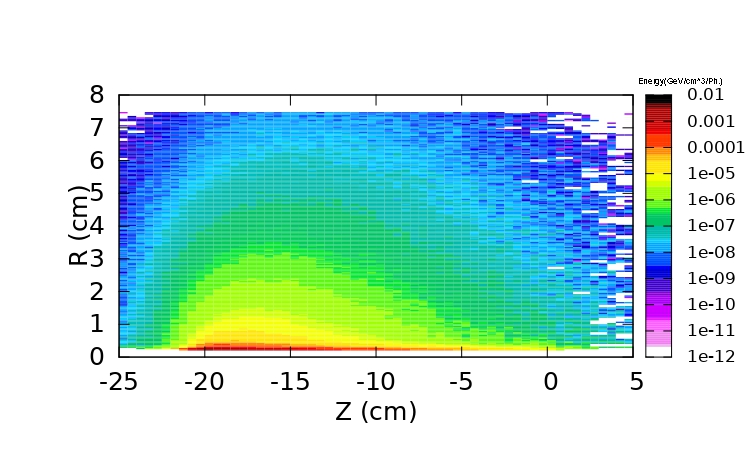}
\caption{The energy deposited along the mask for 350 GeV center-of-mass energy.}
\label{fig:LCWS212D175}
\end{figure}

\begin{figure}[h]
\centering
\includegraphics[scale=1.0]{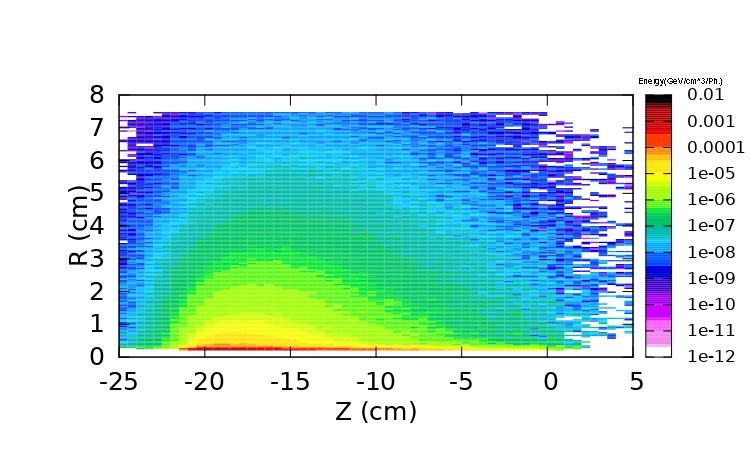}
\caption{The energy deposited along the mask for 500 GeV center-of-mass energy.}
\label{fig:LCWS212D250}
\end{figure}

Figures \ref{fig:LCWS212D175} and \ref{fig:LCWS212D250} show the energy deposited along the mask for both 350 and 500 GeV center-of-mass energy.

Figure \ref{fig:LCWS21_1D175_250} illustrates the peak energy deposited at the mask for both 350 GeV and 500 GeV. In both center-of-mass energies, the peak energy deposited along the mask is at the end of the first tapered part. This means when the mask radius becomes 0.22cm, the peak deposited energies at the mask are 0.0032 and 0.0026 (GeV/$cm^3$/Ph) for both, respectively.

The photon mask can stop $\simeq$ 98.9\% and $\simeq$ 96.1\% of power deposited by both 350 GeV and 500 GeV center-of-mass energies, respectively.

\begin{figure}[h]
\centering
\includegraphics[scale=0.88]{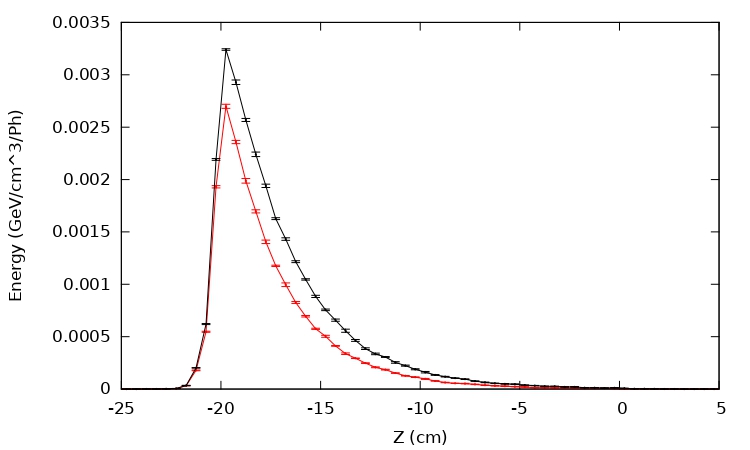}
\caption{The maxmimum energy deposited at the mask for both 350 GeV (Black line) and 500 GeV (Red line).}
\label{fig:LCWS211D175250}
\end{figure}

The Peak Energy Deposited Density (PEDD) can be calculated using the peak energy deposited and the information shown in table \ref{HUSR}. The PEDDs at the mask in case of both 350 and 500 GeV center-of-mass energies are 5.16 and 0.7 J/g/Pulse, respectively. 

The PEDD is helpful to calculate the highest increase in the temperature. The specific heat capacity and density of copper and  PEDDs are used to calculate the maximum temperature rise at the mask for both center-of-mass energies. The maximum temperature rises at the mask for both center-of-mass energies are 13 and 1.8  K/Pulse. In both cases, the tolerable energy density as derived from the endurance strength and tolerable instantaneous temperature rise are below the acceptable limit \cite{maslov2006layout}. Therefore the photon masks are safe against damage.

\section{The Effects of Inserting Photon Masks in the Photon Distribution} 
A helical undulator-based on positron source is foreseen at the ILC. The helical undulator can not only produce the required positrons intensity but it can also produce polarized positrons.
22 photon masks are needed to keep the power deposited at undulator walls below the acceptable limit. Since the mask diameter is smaller than the undulator diameter, photon masks will remove the photons that are further away from the axis. Photons with larger angles always have low energy. So by removing photons with larger angles and low energy, the average photon energy, power and $P_\gamma$ may be affected. Therefore, the effects of adding photon masks in the photon beam at the target is discussed as follows:

\begin{figure}[h]
\centering
\includegraphics[scale=0.50]{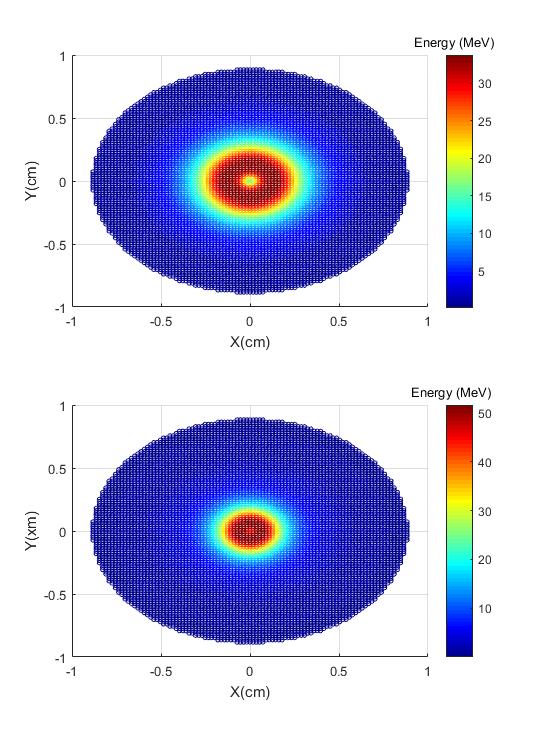}
\caption{Photon distribution at the target plane without masks for both 350 (top) and 500 (bottom) GeV center-of-mass energies.}
\label{fig:LCWS212D175250}
\end{figure}

Figure \ref{fig:LCWS212D175250} shows the photon distribution at the target plane without masks for both 350 and 500 GeV center-of-mass energies, respectively.

\begin{figure}[h]
\centering
\includegraphics[scale=0.53]{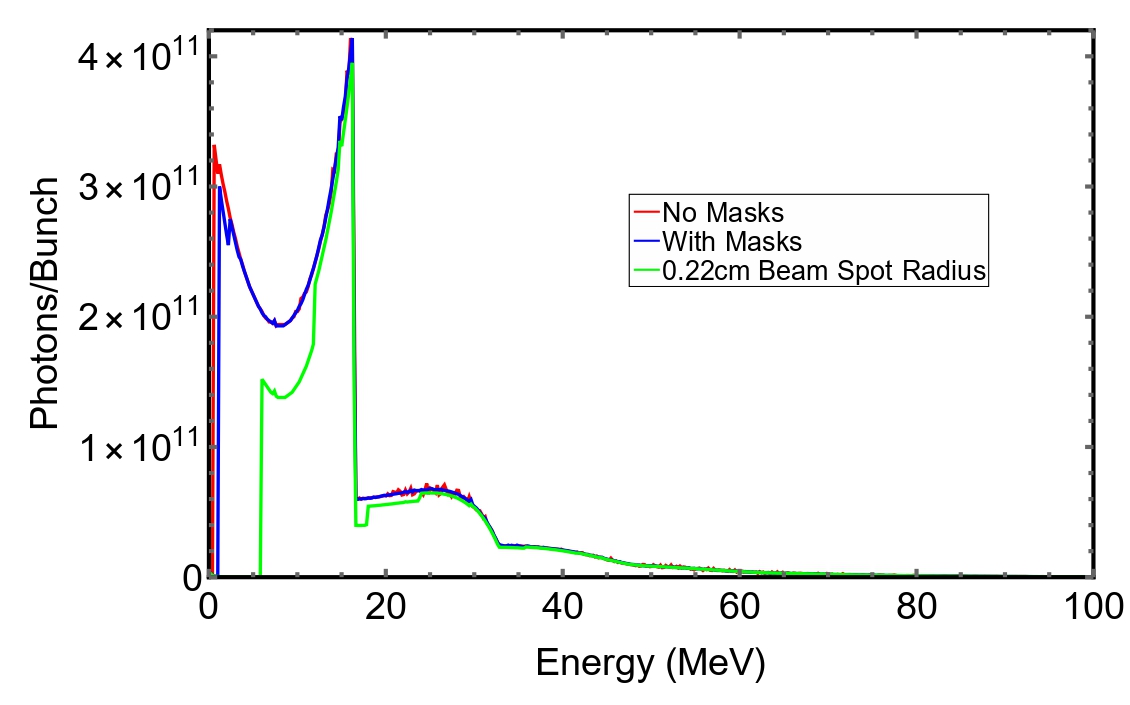}
\caption{The photon spectrum at the target plane for 350 GeV center-of-mass energy with, without photon masks and if only beam spot with 0.22 cm radius is considered.}
\label{fig:LCWS212D350}
\end{figure}

\begin{figure}[h]
\centering
\includegraphics[scale=0.53]{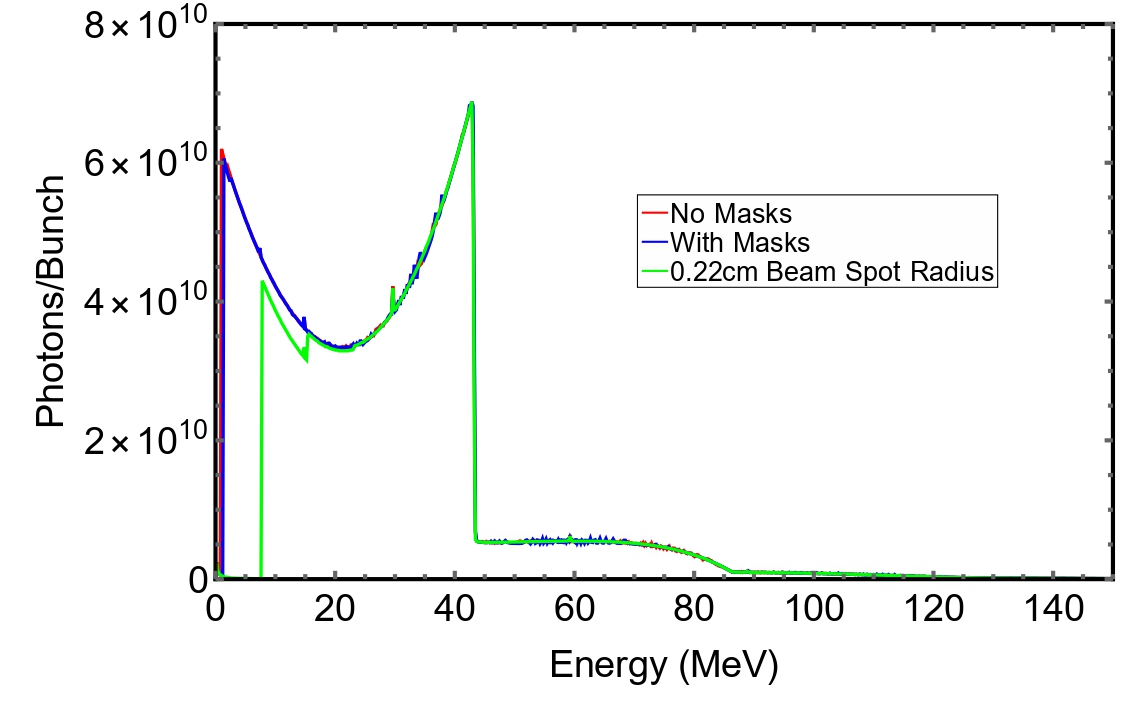}
\caption{The photon spectrum at the target plane for 500 GeV center-of-mass energy with, without photon masks and if only beam spot with 0.22 cm radius is considered.}
\label{fig:LCWS212D500}
\end{figure}

 Figures \ref{fig:LCWS212D350}  and \ref{fig:LCWS212D500} represent the photon spectrum at the target plane with, without photon masks and if only beam with 0.22 cm radius is considered to be 350 and 500 GeV center-of-mass energies, respectively.

\begin{figure}[h]
\centering
\includegraphics[scale=0.57]{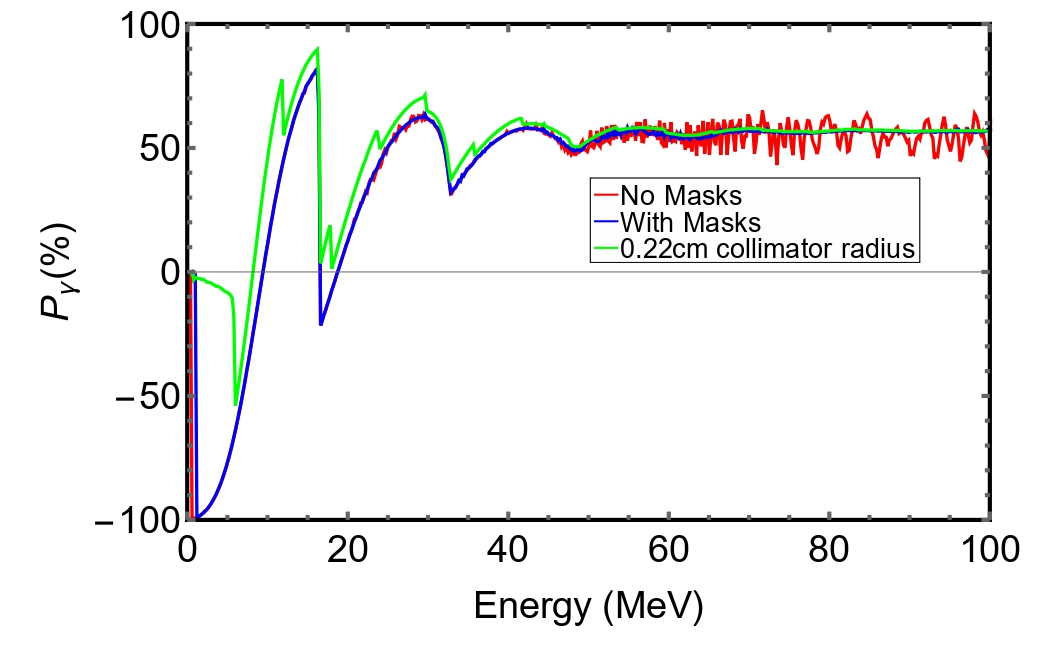}
\caption{The $P_\gamma$ at the target plane for 350 GeV center-of-mass energy with, without photon masks and if only beam spot with 0.22 cm radius is considered.}
\label{fig:LCWS211D175350}
\end{figure}

\begin{figure}[h]
\centering
\includegraphics[scale=0.55]{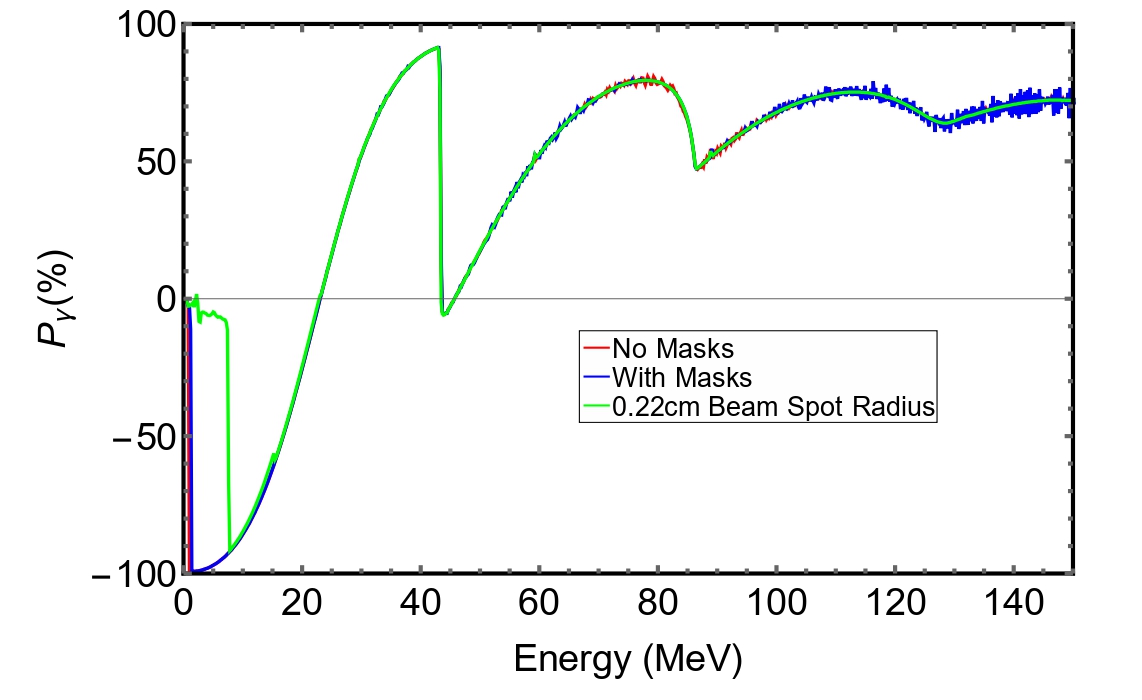}
\caption{The $P_\gamma$ at the target plane for 500 GeV center-of-mass energy with, without photon masks and if only beam spot with 0.22 cm radius is considered.}
\label{fig:LCWS211D175500}
\end{figure}

 Figures \ref{fig:LCWS211D175350} and \ref{fig:LCWS211D175500} show  the $P_\gamma$ at the target plane with, without photon masks and if only beam with 0.22 cm radius is considered for 350 and 500 GeV center-of-mass energies, respectively.

\newpage

Table \ref{tab:FLUKAResultsAtTheTarget} illustrates the effect of adding photon masks on the photon distribution at the target plane for both center-of-mass energies. In addition, the power and average photon energy and $P_\gamma$ of the photon beam with a 0.22 cm radius are shown.

As can be seen, in both center-of-mass energies, the photon masks can increase the average photon energy by $\simeq$ 1 MeV.  In addition, photon masks can increase the $P_\gamma$ from 39 and 44.9 \% to 39.2 and 45.1 \% for both 350 and 500 GeV center-of-mass energies, respectively. 

\begin{table}[h]

	\setlength\tabcolsep{5.0pt}

	\centering
	\caption{Power, average photon energy and $P_\gamma$ at target for both 350 and 500 GeV center-of-mass energies in case of without, with masks and if only beam with 0.22 cm is considered.}
	\label{tab:styles}
	\begin{tabular}{llcc}

		\bottomrule

		{} & { Power (kW)}               & {Energy (MeV)}  & {$P_\gamma$ ($\%$)} \\
\bottomrule
\multicolumn{4}{c}{350 GeV Center-of-mass energy} \\
 \midrule

		{No Masks}  &        60         &     14.16 & 39  \\
		{With Masks}  &    59   &           15.22     &  39.2    \\
              {0.22cm Spot}  &      51    &      19.6           & 56.5    \\

 \midrule
\multicolumn{4}{c}{500 GeV Center-of-mass energy} \\
 \midrule

		{No Masks}  &          43.3       &  27   & 44.9  \\
		{With Masks}  &   43    &            28     &   45.1   \\
              {0.22cm Spot}  &     41.8     &       33          &    49.6  \\

		\bottomrule   %\SI{0.25}{in}
	\end{tabular}
 \label{tab:FLUKAResultsAtTheTarget}
\end{table}

\section{Conclusion}

The power deposited at the photon masks for both 350 and 500 GeV ideal center-of-mass energies have been studied. As expected, 350 GeV center-of-mass energy deposits (186 W) more than the 500 GeV center-of-mass energy (21 W) since the photons in the helical undulator are produced with an opening angle determined by the energy of the electron beam; it is proportional to 1/$\gamma$.

We have modeled a possible photon mask geometry with high absorption efficiency for both center-of-mass energies. Copper mask can stop $\simeq$ 98.9\% and $\simeq$ 96.1\% of power deposited by both 350 GeV and 500 GeV center-of-mass energies, respectively.

PEDDs and maximum temperature increase studies showed that for both center-of-mass energies, the masks are safe. For example, the PEDDs are 5.16 and 0.7 J/(g*pulse), and maximum temperature rises are 13 and 1.8 K/pulse for both 350 GeV and 500 GeV center-of-mass energies, respectively.

Simulation results proved that masks slightly collimate the photon beam. For example, when photon masks are inserted, the $P_\gamma$ can increase from 39 and 44.9 \% to 39.2 and 45.1 \% for both 350 and 500 GeV center-of-mass energies, respectively.

\newpage
\begin{description}
\fontsize{13}{13}\selectfont
\item[$\bullet$ Acknowledgments]
\end{description}
\fontsize{12}{13}\selectfont
GMP was partially supported by the Deutsche Forschungsgemeinschaft (DFG, German Research Foundation) under Germany's Excellence Strategy -  EXC 2121 "Quantum Universe" - 390833306.

\bibliographystyle{plain}
\bibliography{LCWS2021}

\end{document}